\documentclass[preprint,showpacs, showkeys,amssymb,amsmath,a4paper,english]{revtex4}
\usepackage[T1]{fontenc}
\usepackage[latin1]{inputenc}
\usepackage{graphicx}

\makeatletter

\usepackage{babel}
\begin{document}

\title{Filling Gaps in Chaotic Time Series}

\author{Francesco Paparella}

\affiliation{Dipartimento di Matematica {}``E. de Giorgi''\\
Universit\`a di Lecce\\
Lecce, Italy}
\email{francesco.paparella@unile.it}
\pacs{05.45.-a, 05.45.Ac, 05.45.Tp}
\keywords{time series analysis, chaos, filling gaps}

\begin{abstract}
We propose a method for filling arbitrarily wide gaps in deterministic
time series. Crucial to the method is the ability to apply Takens'
theorem in order to reconstruct the dynamics underlying the time
series. We introduce a functional to evaluate how compatible is a
filling sequence of data with the reconstructed dynamics. An algorithm
for minimizing the functional with a reasonable computational effort
is then discussed.
\end{abstract}
\maketitle

\textbf{
One problem faced by many practitioners in the applied sciences is
the presence of gaps (i.e. sequences of missing data) in observed
time series, which makes hard or impossible any analysis. The problem
is routinely solved by interpolation if the gap width is very short,
but it becomes a formidable one if the gap  width is larger than
some time scale characterizing the predictability of the time series. 
}

\textbf{
If the physical system under study is described by a small set of
coupled ordinary differential equations, then a theorem by Takens
\cite{Takens80,Embedology91} suggests that from a single time series
is possible to build-up a mathematical model whose dynamics is
diffeomorph to that of the system under examination.  In this paper we
leverage the dynamic reconstruction theorem of Takens for filling an
arbitrarily wide gap in a time series.
}

\textbf{
It is important to stress that the goal of the method is not that of
recovering a good approximation to the lost data. Sensitive dependence
on initial conditions, and imperfections of the reconstructed
dynamics, make this goal a practical impossibility, except for some
special cases, such as small gap width, or periodic dynamics. We
rather aim at giving one or more surrogate data which can be
considered \emph{compatible} with the observed dynamics, in a sense
which will be made rigorous in the following.
}

\section{\label{sec:Introduction}Introduction}

We shall assume that an observable quantity $s$ is a function of
the state of a continuous-time, low-dimensional dynamical system,
whose time evolution is confined on a strange attractor (that is,
we explicitly discard transient behavior). Both the explicit form
of the equations governing the dynamical system and the function which
links its state to the signal $s(t)$ may be unknown. We also assume
that an instrument samples $s(t)$ at regular intervals of length
$\Delta t$, yelding an ordered set of $\bar{N}$ data 
\begin{equation}
s_{i}=s\left((i-1)\Delta t\right),\,\,\,
i=1,\ldots,\bar{N}.\label{eq:index_i}
\end{equation}
If, for any cause, the instrument is unable to record the value of
$s$ for a number of times, there will be some invalid entries in
the time series $\{ s_{i}\}$, for some values of the index $i$. 

From the time series $\{ s_{i}\}$ we reconstruct the underlying dynamics
with the technique of delay coordinates. That is, we shall invoke
Takens' theorem \cite{Takens80,Embedology91} and claim that the $m$-dimensional
vectors \[
\mathbf{x}_{i}=\left(s_{i},s_{i+\tau},\ldots,s_{i+(m-1)\tau}\right)\]
lie on a curve in $\mathbb{R}^{m}$ which is diffeomorph to the curve
followed in its (unknown) phase space by the state of the dynamical
system which orginated the signal $s(t)$. Here $\tau$ is a positive
integer, and $i$ now runs only up to $N=\bar{N}-(m-1)\tau$. Severals
pitfalls have to be taken into account in order to choose the most
appropriate values for $m$ and $\tau$. Strong constraints also come
from the length of the time series, compared to the characteristic
time scales of the dynamical system, and from the amount of instrumental
noise which affects the data. We shall not review these issues here,
but address the reader to references \cite{Theiler86,NoisyReconstruction91,Provenzale92}.

We note that gaps (that is, invalid entries) in the time series 
$\{s_{i}\}$ do not prevent a successful reconstruction of a set
$\mathcal{R}=\{\mathbf{x}_{i}\}$ of state vectors, unless the total
width of the gaps is comparable with $\bar{N}$. We simply mark as
{}``missing'' any reconstructed vector $\mathbf{x}_{i}$ whose
components are not all valid entries.

If the valid vectors of $\mathcal{R}$ sample well enough the underlying
strange attractor embedded in $\mathbb{R}^{m}$, one may hope to find,
by means of a suitable interpolation technique, a vector field 
$\mathbf{F}:U\rightarrow\mathbb{R}^{m}$
, such that within an open set $U$ of $\mathbb{R}^{m}$ containing
all the vectors $\mathbf{x}_{i}$, the observed dynamics can be approximated
by
\begin{equation}
\dot{\mathbf{x}}=\mathbf{F}(\mathbf{x}).\label{eq:rec_dyn_sys}
\end{equation}
This very idea is at the base of several forecasting schemes, where
one takes the last observed vector $\mathbf{x}_{N}$ as the initial
condition for equation (\ref{eq:rec_dyn_sys}), and integrates it
forward in time (see e.g., \cite{FarmerSidorowich87,Casdagli89}).

The gap-filling problem was framed in terms of forecasts by Serre
\emph{et.  al.} \cite{Serre92}. Their method, which amounts to a
special form of the shooting algorithm for boundary value problems, is
limited by the predictability properties of the dynamics, and cannot
fill gaps of arbitrary width.

The rest of this note is organized as follows: in section
\ref{sec:A-variational-approach} we cast the problem as a variational
one, where a functional measures how well a candidate filling
trajectory agrees with the vector field defining the observed
dynamics. Then an algorithm is proposed for finding a filling
trajectory. In section \ref{sec:An-example} we give an example of what
can be obtained with this method. Finally, we discuss the algorithm
and offer some speculations on future works in
sec. \ref{sec:Discussion}.

\section{\label{sec:A-variational-approach}A variational approach}

The source of all difficulties of gap-filling comes from the following
constraint: the interpolating curve, which shall be as close
as possible to a solution of (\ref{eq:rec_dyn_sys}), must start at the
last valid vector before the gap and reach the first valid vector
after the gap in a time $T$ which is prescribed.

To properly satisfy this constraint,
we propose to frame the problem of filling gaps as a variational one.
We are looking for a differentiable vector function $\xi:[0,T]\rightarrow U$
which minimizes the continuous functional
\begin{equation}
J(\xi)=\int_{0}^{T}\left\Vert
\dot{\xi}(t)-\mathbf{F}\left(\xi(t)\right)\right\Vert
^{2}dt,\label{eq:functional_J}
\end{equation}
with \[\xi(0)=\mathbf{x}_{p};\,\xi(T)=\mathbf{x}_{q}.\]
Defining $l=q-p$, we have $T=l\,\Delta t.$ 
If the curve $\xi(t)$ coincided with the missing curve $\mathbf{x}(t)$
for $t\in[0,T]$, and $\mathbf{F}$ where a perfect reconstruction
of the vector field governing the dynamics of the system, then the
functional would reach its absolute minimum $J=0$. The imperfect
nature of $\mathbf{F}$ suggests that any curve which makes $J$ small
enough can be considered, on the basis of the available information,
a surrogate of the true (missing) curve. In section \ref{sec:An-example}
we shall offer a simple criterion to quantify how small is {}``small
enough''. For the moment we only care to remark that, even for a
perfect reconstruction of the vector field, a curve $\xi$ making
$J$ abitrarily small, but not zero, need not to approximate $\mathbf{x}(t)$,
in fact, the two curves may be quite different; however, such a curve
$\xi$ is consistent with the dynamics prescribed by (\ref{eq:rec_dyn_sys}).

Most methods for minimizing functionals focus on approaching as
quickly as possible the local minimum closest to an initial guess.
Because we may confidently assume that $J$ has many relative minima
far away from zero, we expect that these algorithms will fall on one
of these uninteresting minima for most choices of the initial guess.
Thus, our problem really reduces to that of finding a suitable initial
guess.

The complexity of the problem is greatly limited if we require that
the set $\mathcal{L}=\{\mathbf{y}_{j}\}$,  whose $l+1$ elements
sample the initial guess, has to be a subset of the set $\mathcal{R}$
of reconstructed vectors. The index $j=0,1,\ldots,l$ does not
necessarily follow the temporal order defined in $\mathcal{R}$ by the
index $i$ (cfr. eq. (\ref{eq:index_i})), but we require that
$\mathbf{y}_{0}\equiv\mathbf{x}_{p}$ and
$\mathbf{y}_{l}=\mathbf{x}_{q}$. 
%If we can trust that the set $\mathcal{R}$ of reconstructed vectors
%spans the whole attractor, it seems natural to seek a subset
%$\mathcal{L}=\{\mathbf{y}_{j}\}$ of $\mathcal{R}$, whose $l+1$ elements
%sample a candidate initial guess, with
%$\mathbf{y}_{0}\equiv\mathbf{x}_{p}$ and $\mathbf{y}_{l}=\mathbf{x}_{q}$. 
%The index $j=0,1,\ldots,l$ does not necessarily follow the
%temporal order defined in $\mathcal{R}$ by the index $i$
%(cfr. eq. (\ref{eq:index_i})).  
We shall denote with $S(\mathbf{y}_{j})$ the successor of the vector
$\mathbf{y}_{j}$ with respect to the temporal order in $\mathcal{R}$,
and with $P(\mathbf{y}_{j})$ its predecessor. We propose that a good
choice for the set $\mathcal{L}$ is one which makes as small as
possible the following discretized form of the functional $J$:
\begin{equation}
J_{0}\left(\{\mathbf{y}_{j}\}\right)=\sum_{j=1}^{l-1}\left|\mathbf{y}_{j+1}-\mathbf{y}_{j-1}-S(\mathbf{y}_{j})+P(\mathbf{y}_{j})\right|^{2}.\label{eq:functional_J0}
\end{equation}
Of course, we shall restrict our choice of vectors to be included in
$\mathcal{L}$ only to valid vectors of $\mathcal{R}$ having valid
predecessor and successor. $J_{0}$ could be zero if and only if
$\mathcal{L}$ cointained all the missing vectors in the correct order:
$\mathcal{L}=\{\mathbf{x}_{p},\mathbf{x}_{p+1},\ldots,\mathbf{x}_{q}\}$.
The value of $J_{0}$ increases every time that the order according to
the $j$-index is different from the natural temporal order, that is,
every time that $S(\mathbf{y}_{j})\neq\mathbf{y}_{j+1}$ or
$P(\mathbf{y}_{j})\neq\mathbf{y}_{j-1}$.  If this happens we say that
there is a \emph{jump} in $\mathcal{L}$ between, respectively, the
position $j$ and $j+1$, or $j$ and $j-1$.

Although a set $\mathcal{L}$ which performs many very small jumps
may conceivably attain a very low value of $J_{0}$, there is an exceedingly
small probability to find it within a finite dataset. An intuitive
demonstration of this statement comes from the histogram in figure
\ref{cap:distance-histo}, which shows the distribution of distancies
between each reconstructed vector and its closest neighbor for the
dataset discussed in section \ref{sec:An-example}: as expected the
frequency of closest neighbors quickly drops to zero for short distancies.
Then our strategy will be that of looking for a set $\mathcal{L}$
which performs as few jumps as possible. 

Let us call \emph{orbit} any sequence of valid vectors which does
not jump. The first vector of an orbit shall have a valid predecessor,
and the last a valid successor. Thus we define the \emph{predecessor
of the orbit} as the predecessor of its first vector and likewise
the \emph{successor of the orbit} as the successor of its last vector.
We say that an orbit is \emph{consecutive} to a point if its successor
or its predecessor is the closest neighbor of the point. Two orbits
are \emph{consecutive} if the successor of one orbit is the closest
neighbor of the first vector of the other orbit, or if the predecessor
of one orbit is the closest neighbor of the last vector of the other
orbit. Let us call \emph{branch} a set made of consecutive orbits.
Below we describe a simple algorithm to construct a set $\mathcal{L}$
by joining together one or more consecutive orbits.

\begin{enumerate}
\item We follow forward in time the orbit consecutive to
  $\mathbf{x}_{p}$ for $l$ steps, or until it has a valid
  successor. We store away the set of points made of $\mathbf{x}_{p}$
  followed by the points of this orbit as the $1$\emph{-jump}
  \emph{forward branch}.
\item \label{enu:forward}For each point $\mathbf{y}_{j}$ of each
  $(n-1)$\emph{-jumps} forward branch (where $j=r,2r,\ldots\,\le l_{f}$, 
  $r$ is an arbitrary stride, $l_{f}+1$ is the number of points in the
  forward branch, and $l_{f}\le l$,), we follow forward in time the orbit
  consecutive to $\mathbf{y}_{j}$ for $l-j$ steps, or until it has a
  valid successor. We store away all the points up to $\mathbf{y}_{j}$
  of the current forward branch followed by the points of the
  consecutive orbit as one of the $n$\emph{-jumps} forward branches.
\item We repeat step \ref{enu:forward} for a fixed number $n_{f}$ of
  times.
\item We follow backward in time the orbit consecutive to
  $\mathbf{x}_{q}$ for $l$ steps, or until it has a valid
  predecessor. We store away the points of the consecutive orbit
  followed by $\mathbf{x}_{q}$ as the $1$\emph{-jump backward branch}.
\item \label{enu:backward}For each point $\mathbf{z}_{j}$ of each
  $(n-1)$\emph{-jumps} backward branch (where $j=r,2r,\ldots\,\le l_{b}$,
  $r$ is an arbitrary stride, $l_{b}+1$ is the number of points in the
  backward branch, and  $l_{b}\le l$), we follow backward in time the orbit
  consecutive to  $\mathbf{z}_{j}$ for $j$ steps, or until
  it has a valid predecessor. We store away all the points of this
  orbit followed by all the points from $\mathbf{z}_{j}$ to the end of
  the current backward branch as one of the $n$\emph{-jumps}
  backward branches.
\item We repeat step \ref{enu:backward} a fixed number $n_{b}$ of
  times.
\item For all possible pairs made by one forward branch and one
  backward branch we examine \emph{synchronous} pairs of points, that
  is a point $\mathbf{y}_{j_f}$ in the forward branch, and a point
  $\mathbf{z}_{j_b}$ in the backward branch such that
  $j_f+l_{b}-j_b=l$, where $l_{b}+1$ is the number of points in the
  backward branch. If they coincide, or one is the closest neighbour
  of the other, then we define
  $\mathcal{L}=\{\mathbf{y}_{0},\ldots,\mathbf{y}_{j_f},\mathbf{z}_{j_b+1},\ldots,\mathbf{z}_{l_{b}+1}\}$.
\end{enumerate}
%%%%%%%%%%%%%%%%%%%%%%%%%%%%%%%%%%%%%%%%%%%%%%%%%%%%%%%%%%%%%%%%%%%%%%%%%%

The presence of jumps in $\mathcal{L}$ makes it unsuitable as a
filling set for the gap. Furthermore, the discretized form
(\ref{eq:functional_J0}) of the functional (\ref{eq:functional_J}) is
meaningful only if its arguments $\{\mathbf{y}_{j}\}$ are a subset of
the set of reconstructed vectors $\mathcal{R}$. Then we need a
different discretization of (\ref{eq:functional_J}) which allows as
 argument any point of $U$. The simplest among many possibilities relies on
finite differences, leading to the following expression
\begin{equation}
J_{1}\left(\{\mathbf{w}_{j}\}\right)=\sum_{j=1}^{l}\left[\frac{\mathbf{w}_{j}-\mathbf{w}_{j-1}}{\Delta t}-\mathbf{F}(\mathbf{w}_{j-1/2})\right]^{2}\label{eq:func_J1}
\end{equation} 
where the vectors $\mathbf{w}_{j}$ may or may not belong to
$\mathcal{R}$, and $j=0,1,\ldots,l$.  Here
$\mathbf{F}(\mathbf{w}_{j-1/2})$ is the vector field $\mathbf{F}$
evaluated at the midpoint between $\mathbf{w}_{j-1}$ and
$\mathbf{w}_{j}$.  $J_{1}$ is a function of $m(l-1)$ real variables
($\mathbf{w}_{0}=\mathbf{w}_{p}$ and $\mathbf{w}_{l+1}=\mathbf{w}_{q}$
shall be kept fixed), which can be minimized with standard techniques,
using $\mathcal{L}$ as the initial guess.

%The discretized form (\ref{eq:functional_J0}) of the functional
%(\ref{eq:functional_J}) is meaningful only if its arguments
%$\{\mathbf{y}_{j}\}$ are a subset of the set of reconstructed vectors
%$\mathcal{R}$. To smooth-out the jumps in $\mathcal{L}$ we need a
%different discretization of (\ref{eq:functional_J}) which allows for
%arbitrary arguments. The simplest among many possibilities relies on
%finite differences, leading to the following expression
%\begin{equation}
%J_{1}\left(\{\mathbf{y}_{j}\}\right)=\sum_{j=0}^{l}\left[\frac{\mathbf{y}_{j+1}-\mathbf{y}_{j}}{\Delta t}-\mathbf{F}(\mathbf{y}_{j+1/2})\right]^{2}\label{eq:func_J1}\end{equation}
%where $\mathbf{F}(\mathbf{y}_{j+1/2})$ is the vector field $\mathbf{F}$
%evaluated at the midpoint between $\mathbf{y}_{j}$ and $\mathbf{y}_{j+1}$.
%$J_{1}$ is a function of $lm$ real variables ($\mathbf{y}_{0}=\mathbf{x}_{p}$
%and $\mathbf{y}_{l+1}=\mathbf{x}_{q}$ shall be kept fixed), which
%can be minimized with standard techniques, using $\mathcal{L}$ as
%the initial guess.

\section{An example \label{sec:An-example}}

In this section we show how the algorithm described above performs on
a time series generated by a chaotic attractor. We numerically
integrate the Lorenz equations \cite{Lorenz63} with the usual
parameters ($\sigma=10$, $r=28$, $b=8/3$). We sample the $x-$variable
of the equations with an interval $\Delta t=0.02$, collecting $5000$
consecutive data points which are our time series. One thousand
consecutive data points are then marked as {}``not-valid'', thus
inserting in the time series a gap with a width of 1/5th of the series
length, corresponding to a time $T=20$. For this choice of parameters
the Lorenz attractor has a positive Lyapunov exponent $\lambda \approx
0.9$ \cite{Froyland83}, setting the Lyapunov time scale at
$\lambda^{-1}\approx 1.1$. We also find that the autocorrelation
function of the time series drops to negligible values in about 3 time
units. We conclude that $T$ is well beyond any realistic
predictability time for this time series.

In the present example we selected the embedding delay $\tau=5$
simply by visual inspection of the reconstructed trajectory, and we
choose an embedding dimension $m=3$. However, we checked that results
are just as satisfactory up to (at least) embedding delay $\tau=15$,
and embedding dimension $m=6$.

We apply the algorithm with $n_{f}=2$ and $n_{b}=0$. The strides
are $r=1$ for the 2-jumps forward orbits and $r=100$ for the 3-jumps
forward orbits. This leads to 11001 forward orbits to be compared
with 1 backward orbit, looking for synchronous pairs points which
are neighbour of each other. We find two such pair of points, and the
corresponding two initial guesses $\mathcal{L}_1$ and  $\mathcal{L}_2$
are such that $J_{0}(\mathcal{L}_1)=1.21$ and
$J_{0}(\mathcal{L}_2)=1.04$. 

The approximating vector field $\mathbf{F}$ is extremely simple, and
its choice is dictated solely by ease of implementation. A comparison
between different interpolating techniques is off the scope of this
paper, and the interested reader may find further information in
\cite{Casdagli89}.  If $\bar{\mathbf{x}}_{j}$ and
$\bar{\bar{\mathbf{x}}}_{j}$ are the vectors of $\mathcal{R}$,
respectively, closest and second closest to $\mathbf{w}_{j}$, then we
define
\begin{equation}
\mathbf{F}(\mathbf{w}_{j-1/2})=\frac{\bar{\mathbf{x}}_{j}-P(\bar{\mathbf{x}}_{j})+\bar{\bar{\mathbf{x}}}_{j}-P(\bar{\bar{\mathbf{x}}}_{j})}{2\Delta t}.\label{eq:my-interpol}
\end{equation}
Using the definition (\ref{eq:my-interpol}) in (\ref{eq:func_J1}),
we obtain $J_{1}(\mathcal{L}_1)=3.46$ and $J_{1}(\mathcal{L}_2)=3.17$.
In order to smooth-out the jumps in the filling sets, the function
$J_{1}$ is further decreased by iterating five times a steepest-descent
line minimization (see, e.g., \cite{NumRecipesII}) using $\mathcal{L}_1$
and $\mathcal{L}_2$ as initial guesses. This procedure yelds two sets
of $l+1$ points, $\mathcal{M}_1$ and $\mathcal{M}_2$ such that
$J_{1}(\mathcal{M}_1)=1.54$,  and $J_{1}(\mathcal{M}_2)=1.34$. The
corresponding time series are shown in figure \ref{cap:fillings}.
The difference between the smoothed sets $\mathcal{M}_1$ and
$\mathcal{M}_2$ plotted in figure \ref{cap:fillings} and the sets with
jumps $\mathcal{L}_1$ and $\mathcal{L}_2$ would be barely noticeable
on the scale of the plot. 

The effect of the smoothing may be appreciated by looking at figure
\ref{cap:smoothing} which shows the region across the jump between
two consecutive orbits of $\mathcal{L}_1$. The non-smoothed
filling set (dashed line) abruptly jumps from one orbit to the other,
but the smoothed trajectory (thick solid line) singled out by the
points of $\mathcal{M}_1$ gently moves between them. 

No attempt has been made to approach as closely as possible the local
minimum of $J_{1}$. In fact, we verified that for orbits in
$\mathcal{R}$ having the same length as the interpolating sets
$\mathcal{M}_1$ and $\mathcal{M}_2$, $J_{1}$ ranges (roughly) between
1 and 9. This is a measure of the accuracy with wich the field
$\mathbf{F}$ approximates the true dynamics of the observed system,
and there is no point in looking for an interpolating set having a
value of $J_{1}$ below this range.

\section{\label{sec:Discussion}Discussion and conclusions.}

In this note we have described an algorithm which fills an arbitrarily
wide gap in a time series, provided that the dynamic reconstruction
method of Takens is applicable. The goal is to provide a filling signal
which is consistent with the observed dynamics, in the sense that,
in the reconstructed phase space, the vector tangent to the filling
curve should be close to the vector field modeling the observed dynamics.
This request is cast as a variational problem, defined by the functional
(\ref{eq:functional_J}). The acceptable degree of closeness is determined
by the level of accuracy of the reconstruction, which we quantify
by computing the discretized form (\ref{eq:func_J1}) of the functional
for orbits having the same length of the gap.

Obviously, if the time series has more than one gap, our method can
be applied to all the gaps, indipendently from each other.

A second novel idea, that greatly simplyfies the problem, is that of
building a rough solution by stitching together pieces of the observed
dataset. The actual solution, which will not be an exact copy of
anything present in the observed dataset, is obtained by refining this
first draft. We have illustrated a basic algorithm that embodies this
idea, although no attempt has been made at making it computationally
optimal. In particular, with the algorithm in its present form, many
of the forward and backward branches will be partial copies of each
other, because nothing forbids two distinct branches to jump on the
same orbit. This leaves some room for improvement, because the
effectiveness of the method relies on a 
substantial amount of the set $\mathcal{R}$ of reconstructed vectors
to be explored by a relatively limited number of branches.  In a
forthcoming, enhanced version of the algorithm some kind of tagging
mechanism shall be incorporated in order to produce non-overlapping
hierarchies of forward and backward branches.

We observe that this algorithm does not give a guarantee of success:
it is perfectly possible that no point of the forward branches is
the closest neighbor of (or coincides with) a synchronous point of
the backward branches. In this case the obvious attempt is to deepen
the hierarchy of the branches, as much as it is computationally feasible.
Or, one may relax the request that branches may jump only between
closest neighbors, and accept jumps between second or third neighbors
as well. As a last resort, one may stitch any pair of forward and
backward branches at their closest synchronous points, hoping that
the resulting jump could later be smoothed satisfactorily by minimizing
the function (\ref{eq:func_J1}). However, when facing a failure of
the algorithm, we believe that first should be questioned the goodness
and appropriateness of the dynamic reconstruction. The presence of
too many gaps, the shortness of the time series, or measurement inaccuracies
may make the gap-filling problem an insoluble one. We speculate that
the ability of filling gaps with relative ease is a a way to test
the goodness of a dynamic reconstruction. 

The ease with which a gap may be filled, as a function of his width,
is a problem deserving further work. For the moment we simply recall
that if a set of initial conditions of non-zero measure is evolved
in time according to (\ref{eq:rec_dyn_sys}), eventually we expect
it to spread everywhere on the attractor (here the measure is the
physical measure $\mu$ of the attractor cfr. ref. \cite{EkRue85}).
More rigorously, if $\phi_{t}$ is the flow associated to (\ref{eq:rec_dyn_sys}),
and if it is a mixing transformation, then, for any pair of sets $A$,
$B$ of non-zero measure, $\lim_{t\rightarrow\infty}\mu(\phi_{t}A\cap
B)=\mu(A)\mu(B)$. 
The dispersion of a set of initial conditions is further discussed
in \cite{Sklar73}, where, for example, it is shown that the essential
diameter of a set of initial conditions cannot decrease in time, after
an initial transient of finite length. 

This leads to the idea that wide gaps should be easier to fill than
not-so-wide ones, because forward and backward branches have explored
larger portions of the attractor, and so there is a greater chance
to find synchronous points where they can be joined together. As a
first step toward the verification of this hypotesis, we computed
the average minimum distance between synchronous points of the 1-jump
forward and backward branch as the gap moves along the dataset, for
several gap widths. We used the dataset discussed in sec. \ref{sec:An-example}
and a ten times longer extension of it. The results, plotted in figure
\ref{cap:gap-vs-dist}, show that the average separation of the branches
initially increases as the gap widens, but then it reaches a well-defined
maximum and, from there on, decreases as the gap width is further
increased. 

We close by mentioning that the dynamic reconstruction technique has
been successfully applied even to stationary stochastic time series,
to generate surrogate data with the same statistics of the observed
ones \cite{RAP97}. This fact, and the hypothesis that ergodicity
(or the stronger requirement of being mixing), rather than determinism,
is the crucial property that allows for filling gaps, suggest that
some modified version of our method should be able to fill gaps in
a large class of stochastic time series.

\section*{Acknowledgements}
This work has been supported by  \emph{fondo convezione strana} of the
Department of Mathematics of the University of Lecce. We are grateful to
Prof. Carlo Sempi for valuable comments.

~

~
\newpage
\section*{Figure Captions}
\begin{figure}
\includegraphics{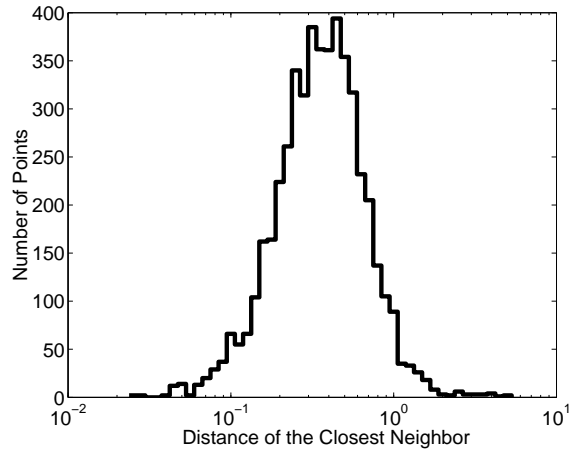}

\caption{Distibution of the distances between each reconstructed vector and
its closest neighbor for the dataset discussed in sec. \ref{sec:An-example}.
\label{cap:distance-histo}}
\end{figure}

\textbf{Figure \ref{cap:distance-histo}}
Distibution of the distances between each reconstructed vector and
its closest neighbor for the dataset discussed in sec. \ref{sec:An-example}.

\begin{figure}
\includegraphics{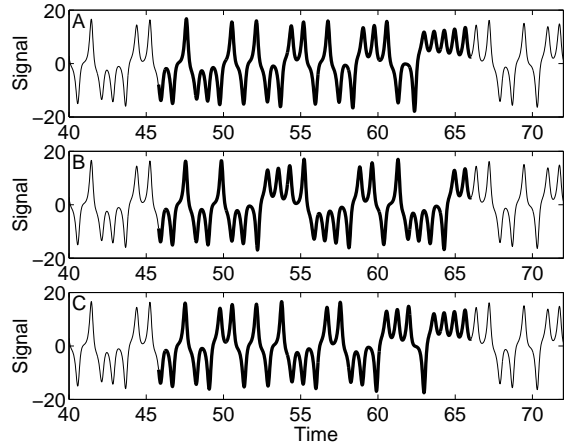}

\caption{Panel A) shows portion of the time series discussed in sec. \ref{sec:An-example}.
The blackened line was removed and the resulting gap was filled by
applying the algorithm described in section \ref{sec:A-variational-approach}.
The blackened lines in panels B) and C) are two different fillings.
\label{cap:fillings}}
\end{figure}

\textbf{Figure \ref{cap:fillings}}
Panel A) shows portion of the time series discussed in
sec. \ref{sec:An-example}.  The blackened line was removed and the
resulting gap was filled by applying the algorithm described in
section \ref{sec:A-variational-approach}.  The blackened lines in
panels B) and C) are two different fillings.

\begin{figure}
\includegraphics{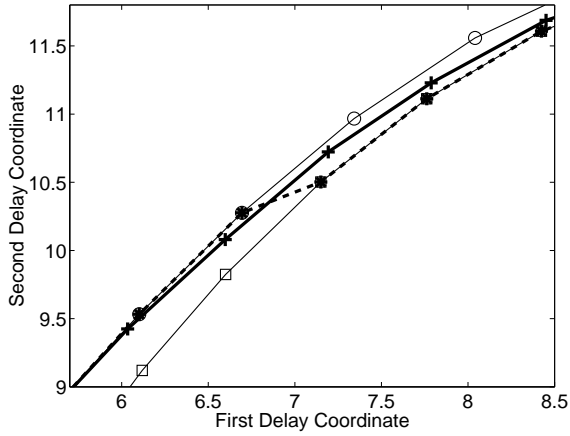}

\caption{Plot of the first two components of the reconstructed vectors of:
$\mathcal{M}_1$ (thick solid line with crosses); $\mathcal{L}_1$ 
(thick dashed line with asterisks); one orbit of $\mathcal{L}_1$ and 
its successors (thin line with open circles); the orbit consecutive to
it and its predecessors (thin line with open squares). To illustrate
the smoothing effect of minimizing functional (\ref{eq:func_J1}), we
only plot a very small portion of these sets in the vicinity of the
jump between the consecutive orbits. \label{cap:smoothing}}
\end{figure}

\textbf{Figure \ref{cap:smoothing}}
Plot of the first two components of the reconstructed vectors of:
$\mathcal{M}_1$ (thick solid line with crosses); $\mathcal{L}_1$ 
(thick dashed line with asterisks); one orbit of $\mathcal{L}_1$ and 
its successors (thin line with open circles); the orbit consecutive to
it and its predecessors (thin line with open squares). To illustrate
the smoothing effect of minimizing functional (\ref{eq:func_J1}), we
only plot a very small portion of these sets in the vicinity of the
jump between the consecutive orbits.

\begin{figure}
\includegraphics{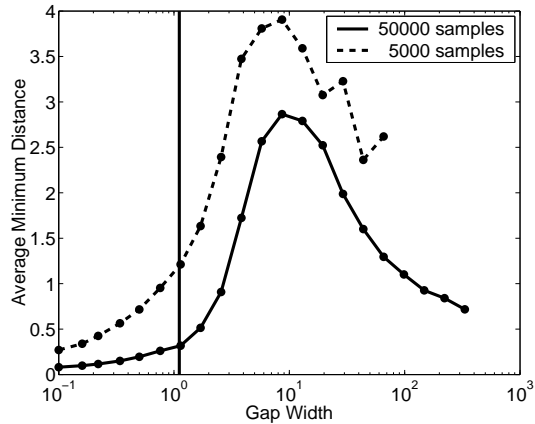}

\caption{Average minimum distance of sinchronous points in the 1-jump forward
and backward branch as a function of gap width. The dashed line refers
to the dataset discussed in sec. \ref{sec:An-example}, the solid
line refers to a ten times longer extension of that dataset. The
vertical line marks the Lyapunov time
$\lambda^{-1}\approx1.1$. \label{cap:gap-vs-dist}} 
\end{figure}

\textbf{Figure \ref{cap:gap-vs-dist}}
Average minimum distance of sinchronous points in the 1-jump forward
and backward branch as a function of gap width. The dashed line refers
to the dataset discussed in sec. \ref{sec:An-example}, the solid
line refers to a ten times longer extension of that dataset. The
vertical line marks the Lyapunov time
$\lambda^{-1}\approx1.1$.

\newpage

\end{document}